\documentclass{article}
\usepackage[utf8]{inputenc}
\usepackage{authblk}
\usepackage{graphicx}
\usepackage{tabularx}
\usepackage{caption}
\usepackage{subcaption}
\usepackage{hyperref}
\usepackage{amssymb}
\usepackage{pifont}
\newcommand{\cmark}{\ding{51}}%
\newcommand{\xmark}{\ding{55}}%
\usepackage{array}
\usepackage{dblfloatfix}
\usepackage{ulem}
\newcolumntype{P}[1]{>{\centering\arraybackslash}p{#1}}

\title{Quantifying the Privacy-Utility Trade-offs in COVID-19 Contact Tracing Apps}
\author[1]{Patrick Ocheja}

\author[1]{Yang Cao}
\author[1]{Shiyao Ding}
\author[1]{Masatoshi Yoshikawa}
\affil[1]{Graduate School of Informatics \\ Kyoto University}

\begin{document}

\maketitle

\section{Introduction}
COVID-19 is a novel severe acute respiratory syndrome coronavirus 2 (SARS-CoV-2) that was first reported in late December 2019 in the city of Wuhan, China \cite{xu2020pathological}. The virus can be transmitted from person to person which could occur through physical contact with body fluid from an infected person or by touching infected objects or surfaces and then touching one's face, mouth or nose. Since the first case of COVID-19, over four million persons have been diagnosed with COVID-19 with more than three thousand fatalities reported across the world \cite{dong2020interactive}. 

To combat the virus, the World Health Organization (WHO) recommends strategic guidelines such as limiting human-to-human transmissions, early identification, isolation and care for infected persons, and communicating critical risk and event information to communities so as to avoid misinformation \cite{whocovid202012}. Consequently, various countries have adopted measures such as contact tracing, travel restrictions, and partial or total shutdown of economic activities so as to flatten the curve.

To effectively implement the recommendations for containing the spread of COVID-19, it is necessary for government agencies, healthcare organizations and citizens to be aware of vital information such as infectious clusters, likelihood of infection, and personal status. This will require access to personal data such as healthcare data, location information, and recent contacts. Therefore, it also becomes important to protect the privacy rights of citizens as their personal data are being collected.


This work presents a quick review of some of the various methods that have been adopted in combating and controlling the spread of COVID-19 and how privacy rights of citizens are being upheld or infringed. We begin by grouping them according to: the technology used, architecture, trade-offs between privacy and utility, phase of adoption (1, 2 or 3) and notable representative applications.

\section{Related work}
\cite{tang2020privacy,raskar2020apps,cho2020contact} reviewed COVID-19 contact tracing related solutions with the aim of identifying utility/security requirements, their (dis)advantages and present key findings to guide the development of more effective privacy-aware contact tracing solutions. Also, \cite{raskar2020apps,cho2020contact} presented a review of contact tracing tools based on privacy-utility trade-offs and the type of technology  adopted such as  broadcast, selected cast, unicast, participatory and safe paths. 

The unique contribution of our work is to provide a broad classification of tools used in combating the COVID-19 pandemic focusing on distinguishing criteria such as the type of technology used, the architecture of the proposed tool, the phase in which the tool is adopted or useful, and applications backed by representative organizations. This work further assess the implications of total-privacy, privacy trade-offs, and zero-privacy scenarios. Specifically, we present a novel approach for evaluating privacy using both qualitative and quantitative measures of privacy-utility assessment of contact tracing applications. In this new method, we classify utility at three (3) distinct levels: no privacy, 100\% privacy and at \textit{k} where \textit{k} is set by the system providing the utility or privacy. Finally, we make a case for potential solutions that will be useful as countries begin to lift \textit{social distancing} \cite{wilder2020isolation} measures.

\section{Classification of Reviews}
In this section, we review works under 3 broad categories: contact tracing technology, system architecture, and privacy vs utility trade-offs.

\subsection{Contact tracing technology}
In this subsection, we take a classification by position/contact collection technology, which is classified into three types: base station, GPS and bluetooth.\\

\noindent\textbf{Base station}: Telecommunication operators provide anonymous cell phone base station data to the government, which is mainly used to predict and monitor the overall trend of the epidemic. The data  based on the location of the base station can relatively accurately determine the user's mobility, but due to the accuracy insufficient, it can only be used to help judge close contacts.

There some telecommunication companies to provide such data for the government in Europe such as \textit{Telecom Italia} in Italia, \textit{Swisscom} in Swiss and \textit{Deutsche Telekom} in Germany.

\noindent\textbf{GPS}: GPS (Global Positioning System) is a satellite-based radio navigation system owned by the United States government and operated by the United States Space Force. This can obtain precise data of personal local position and achieve the recognition of close contacts to a certain extent and many countries have applied such type of applications.

In Israel, the Israeli Ministry of Health has publish an application called \textit{The Shield} \cite{israle}: the Ministry of Health, when encountered with a new Coronavirus patient, will interview them about all the locations and places they visited before coming there. Once the patient has told this information, it is added onto a JSON file and this file is hourly downloaded onto the app.  So, every hour new locations of possible Coronavirus threats are added onto the app without fail.
While a user is using the app, the software of The Shield will analyze the surroundings and compare it with the JSON file data. If the location of a Coronavirus patient is pinged, then the app will send an alert to tell the user that a match has been found and they are at risk of exposure to COVID-19. On the same hand, if the location is safe, the app will alert you then as well to give you an 'All Clear' sign.
\begin{figure}[htbp]
\includegraphics[scale=0.23,keepaspectratio]{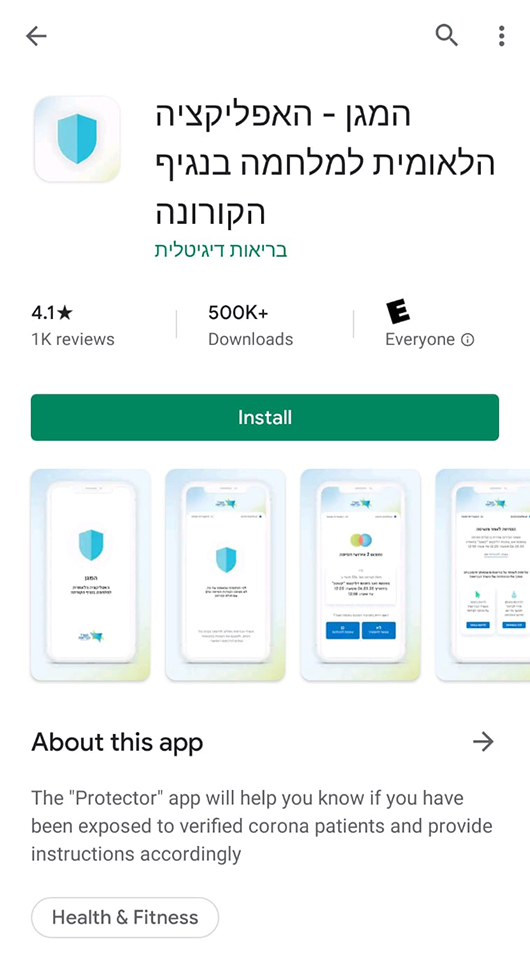}
\centering
\caption{ The interface of 'The Shield' application}
\end{figure}
They have do some safety measures to decrease thereat to the privacy: 1) it guarantees that the information is only stored locally; 2) they have released the source code of the app under an MIT license; 3) they have asked the network security company to conduct a security review of the code; 4) the information of the user is only used if the user specifically volunteers it.

In Korea, the government utilize a system called \textit{Epidemic Intelligent Management System} rather than applications. It only allows a few officials of the epidemic prevention department to log in, and only records patient information "those that may cause super spread"; 
And it need to ask for the agreements of the related companies to protect patient privacy when obtaining official patient information\\

\noindent\textbf{Bluetooth}: bluetooth technology is a wireless technology standard used for exchanging data between fixed and mobile devices over short distances. The applications based on bluetooth are opposite to other technologies where personal invasion is relatively small and it does not exist and actual location. Many countries have applied such type of applications.

In Singapore, the Singaporean government released a \textit{TraceTogether} \cite{trace} application that allows for digital contact tracing using the custom BlueTrace protocol. The user's Bluetooth proximity data is encrypted and stored only on his/her phone. Then, the Ministry of Health (MOH) will seek user's consent to upload the data, if it's needed for contact tracing.
If a user had close contact with a COVID-19 case, TraceTogether allows the MOH call the user more quickly, to provide guidance and care.

In Australia, they published an application called \textit{COVIDSafe} \cite{australia} based on bluetooth. When the user choose to download it, the user will receive a confirmation SMS text message to complete installation. Then the system creates a unique encrypted reference code just for the user and it recognizes other devices with the COVIDSafe app installed and Bluetooth enabled. When the app recognizes another user, it notes the date, time, distance and duration of the contact and the other user's reference code. The COVIDSafe app does not collect your location.

In America, the companies Apple and Google jointly announced that they will release an application about \textit{Health Code} \cite{ChaturvediTop10} using Bluetooth Low Energy (BLE) technology to help governments and public health agencies slow the spread of the epidemic.

In China, the government released an application called as \textit{StayHomeSafe} \cite{li2020global} based on a combination of many location technologies such as bluetooth, Wifi and GPS. After a user download the app, he/she must keep the StayHomeSafe app running on your smartphone, and turn on the Bluetooth, Wi-Fi, Location Services, Camera, Notifications and Mobile Data functions throughout the quarantine period. Then, the app will utilize those informations to judge whether the user is at home or where he/she has been gone.

\subsection{Classification by System Architecture}
COVID-19 applications can be classified based on the architecture of the underlying system in three categories: centralized, distributed and decentralized. Centralized applications are applications whose data is solely managed and regulated by a central authority. As for distributed applications, the data and processing activities could take place across multiple nodes but with a centralized control. In a simple sense, decentralized applications are not regulated or managed by a central authority instead, each user on the application maybe responsible for managing their data and behaviour on the network. However, some decentralized systems maybe centralized in nature depending on the underlying algorithm and how users are admitted. In this work, we consider decentralized systems as a subset of distributed systems.

\begin{figure}[htbp]
\includegraphics[width=1\textwidth,keepaspectratio]{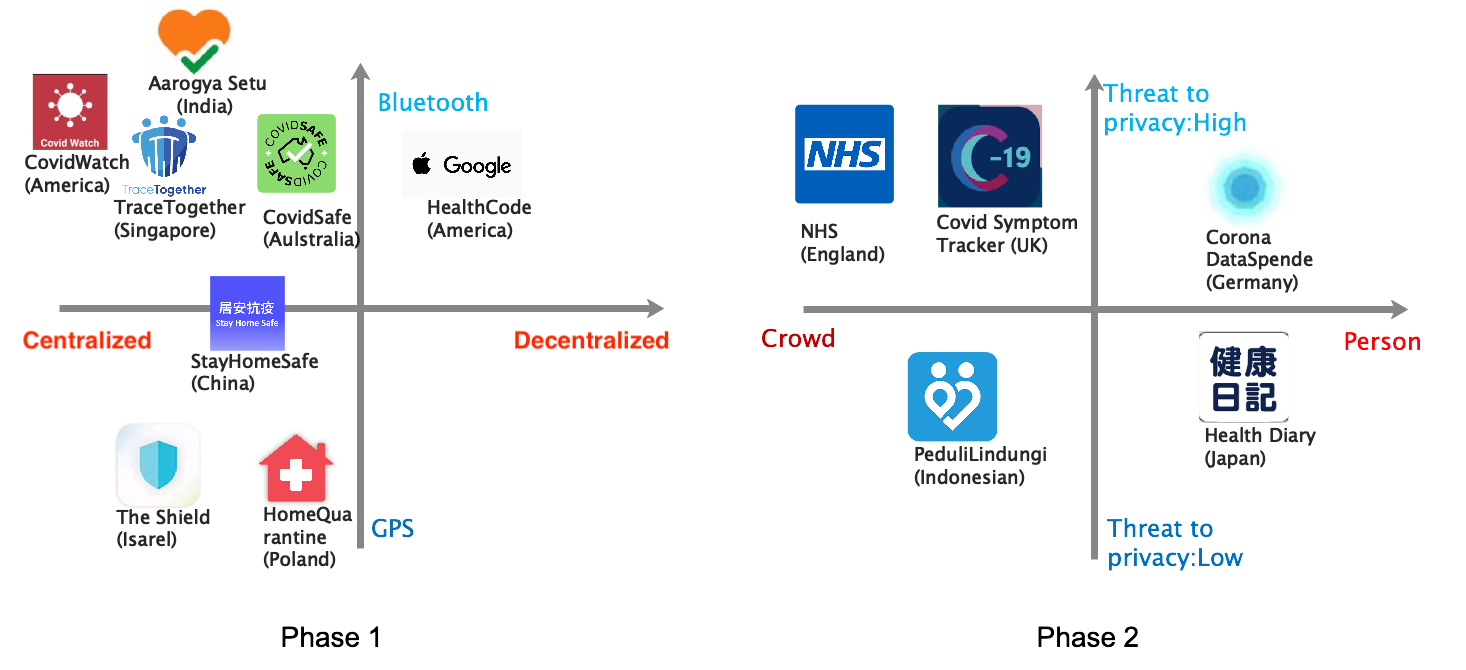}
\caption{A broad classifications based on technologies, architectures and privacy threats}
\end{figure}

In America, the app called \textit{CovidWatch} \cite{CovidWatch2020} is developed in collaboration with Stanford University, the app empowers people to protect themselves and their communities without the need of surrendering their privacy.  It uses Bluetooth signals to detect users when they are in proximity to each other and alerts them anonymously if they were in contact with someone who has tested positive. A distinguishing feature of the app is that any third party, including the government won't be able to track who was exposed by whom. It has been among the first apps to release an open-source protocol for privacy-preserving, decentralized Bluetooth contact tracing.

In Germany, a smartwatch app called \textit{Corona DataSpende} \cite{ChaturvediTop10} is developed that can monitor the spread of coronavirus by collecting crucial signs-pulse rate, body temperature, sleep patterns from volunteers wearing a smartwatch or a fitness tracker. It checks whether they have developed any Covid-19 symptoms or not. The results are then portrayed on an online interactive map that makes it possible for health authorities to take stock of the situation and find out the hotspots.

In India, the Indian Ministry of Electronics and IT developed an app called \textit {Aarogya Setu} \cite{jhunjhunwalarole} which can notify users if they have crossed paths with someone who has been diagnosed positive. Tracking is done via Bluetooth and a location-generated graph that charts proximity with anyone infected. Once the app is installed, the users are required to switch on their Bluetooth and Location sharing, and keep them on always for effective tracking. The app also has a self-testing function. A user has to answer a few questions, and if the responses indicate symptoms of coronavirus then the information is sent to the government servers.

\begin{figure}[htbp]
\includegraphics[scale=0.18,keepaspectratio]{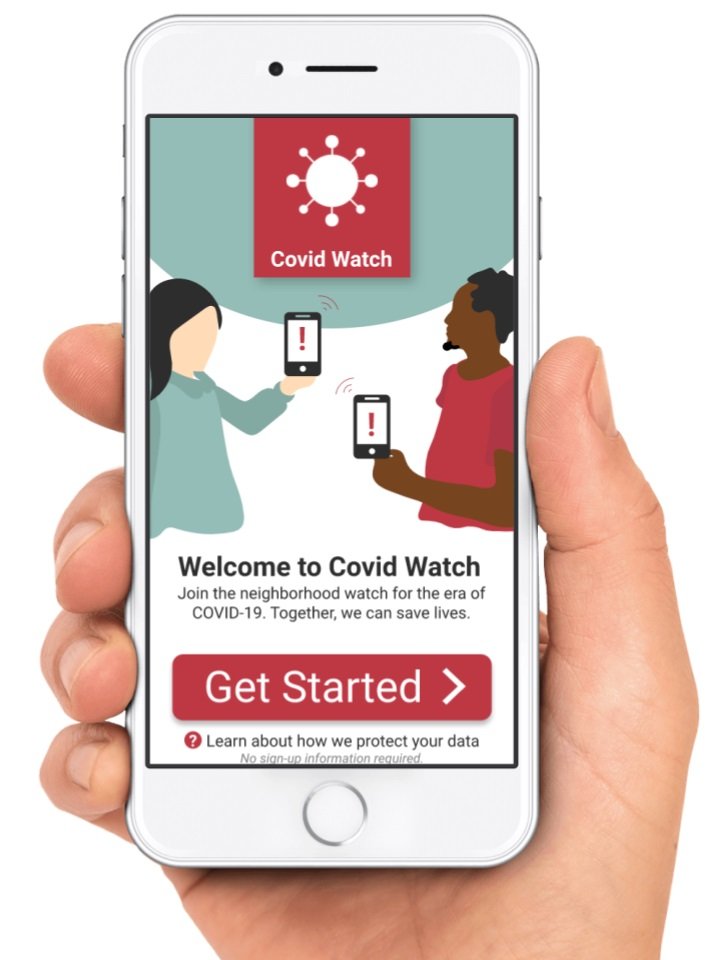}
\includegraphics[scale=0.5,keepaspectratio]{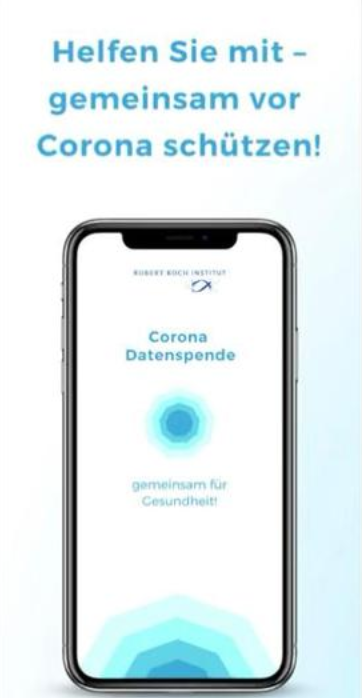}
\includegraphics[scale=0.6,keepaspectratio]{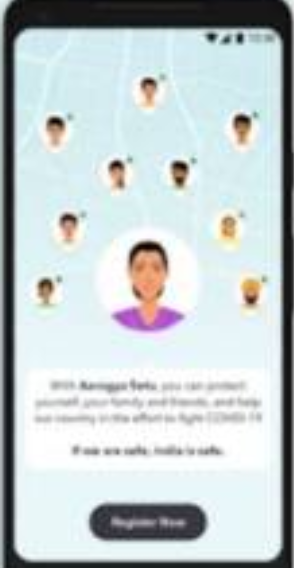}
\caption{1) Covidwatch; 2) Corona DataSpende; 3)AarogyaSetu}
\end{figure}

In UK, an app called \textit{Covid Symptom Tracker} \cite{mayor2020covid} has been designed by doctors and researchers at the King's College London and St. Thomas hospitals, in partnership with a private healthcare company called Zoe Global. The app studies the symptoms of the virus for advanced research and also helps track how it spreads. The scientists analyze high-risk areas in the UK, speed of virus spread, and the most vulnerable group, based on health conditions. The app is GDPR \cite{voigt2017eu} compliant and data is used only for healthcare research and not commercial purposes. 

In England, the \textit{NHS}( National Health Service) developed an application to keep a tab on people's movements and notify those who come in close contact with those who have been infected. Experts suggest that by analysing the patterns of the virus spread and hotspots, the app would also help in relaxing lockdown. It would categorize details based on demography, household structures and mobility patterns and based on this, maximum number of people would be allowed to move freely.

In Indonesian, an app called \textit{PeduliLindungi } \cite{ChaturvediTop10}is
developed by the Indonesian Communications and Information Ministry, along with the State-Owned Enterprises (SOEs) Ministry, the app enables users to compile data related to the spread of COVID-19 in their communities and help boost the government's efforts to track confirmed cases, as well as those suspected to be infected with the virus. It cross references data stored on mobile device through Bluetooth. When a user is in the vicinity of another user whose data has been uploaded to PeduliLindungi, the app enables an anonymous exchange of identities, according to its official website.
\begin{figure}[htbp]
\includegraphics[scale=0.54,keepaspectratio]{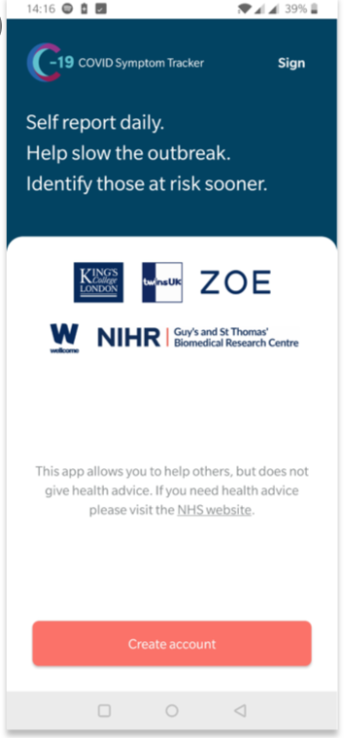}
\includegraphics[scale=0.5,keepaspectratio]{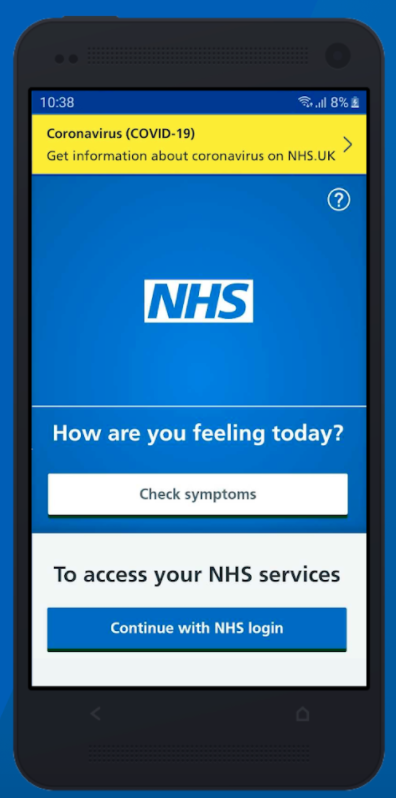}
\includegraphics[scale=0.6,keepaspectratio]{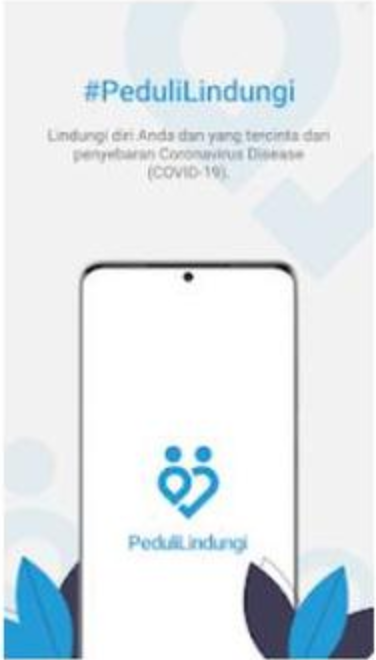}
\caption{1) Covid Symptom Tracker; 2) NHS; 3) PeduliLindungi}
\end{figure}

In Poland, the government has been one of the first Western countries to roll out a smartphone app called \textit{Home Quarantine} \cite{pinkas2020public} that collects a lot of personal information, including people's location and digital photos, in its fight to combat the pandemic. In this app, people upload their selfies when asked by the officials, so that their exact location can be pinpointed. It has become mandatory for anyone who has developed coronavirus symptoms. 

In Japan, a health management application called \textit{Health Diary} is developed by Health Tech Research Institute.
It allows each person to register and manage the check items when a new coronavirus infection (COVID-19) is suspected.
Data is managed only on the smartphone where it will not be sent to the outside including our company unless the person himself sends to the outside based on his intention.
\begin{figure}[htbp]
\includegraphics[width=0.5\textwidth,keepaspectratio]{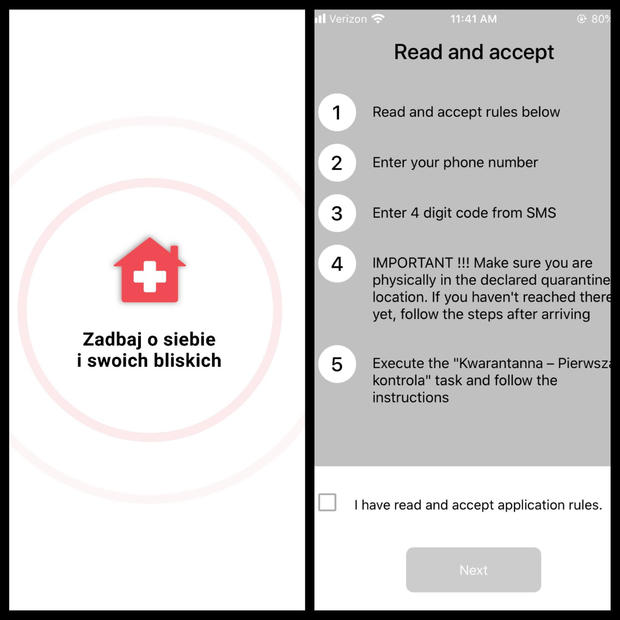}
\includegraphics[width=0.5\textwidth,keepaspectratio]{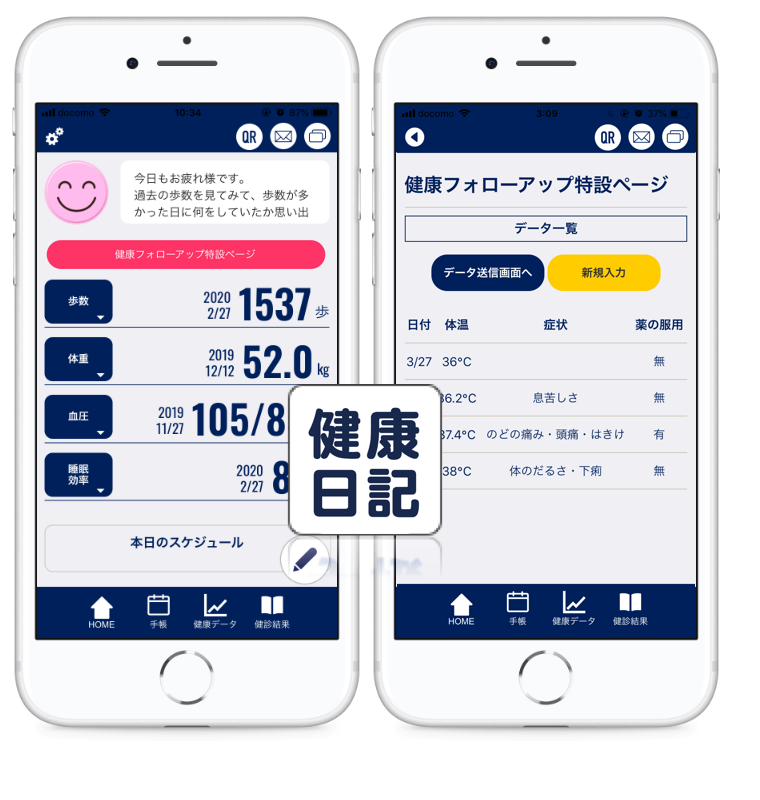}
\caption{1) Home Quarantine; 2) Health Diary}
\end{figure}

\begin{table}
\caption{Decentralized COVID-19 applications.}\label{decentralized_apps}
\begin{tabular}{|p{0.18\linewidth}|p{0.56\linewidth}|p{0.16\linewidth}|}
\hline
\textbf{Application} &  \textbf{Features} & \textbf{Availability}\\
\hline
Safe Places \cite{raskar2020apps} &  Interview verified cases, discover and trace likely contacts with verified cases, provide public anonymized and aggregated data sets of possible virus spread hotspots & Public\\
\hline
COVID SafePaths \cite{raskar2020apps} &  Logs diagnosed patients' history at a 5-minute interval, assist public health officials with such information for contact tracing, and inform users of possible exposure & Public\\
\hline
MiPasa \cite{mipasa2020} &  Swift and precise early detection of carriers and infection hotspots, privacy-based information sharing among stakeholders, and analytics tools for deriving insights. & Public\\
\hline
TNC \cite{tnc2020} &  Privacy-first contact tracing protocol, enables interoperability between contact tracing applications, bluetooth-base communication for reporting and detecting carriers & Public Protocol\\
\hline
PACT \cite{pact2020} &  Broadcasts constantly-changing, anonymous and randomly chosen chirp values, a tracing layer for users to determine their risk to contracting the virus and interact with health workers & Public Protocol\\
\hline
Alipay Blockchain \cite{alipay2020} &  Review, record and track demand, supplies and logistics of epidemic prevention materials. & Private\\
\hline
PHBC \cite{nguyen2020blockchain} &  Track and identify virus-free zones, connect healthcare across organizations, medical supplies alert system, and facilitate joint research. & N/A\\
\hline
\end{tabular}
\end{table}

\subsubsection{More on Decentralized Systems}
In table \ref{decentralized_apps} we show COVID-19 applications that are built on a decentralized architecture.
MIT Safe Paths \cite{raskar2020apps} is a global movement for a free, open-source, privacy-by design tools to help flatten the curve of COVID-19. Two main applications have been developed based on MIT Safe Paths these are: Safe Places and COVID Safe Paths. Safe Places is a browser-based map tool that enables public health officials interview verified cases, discover and trace likely contacts with verified cases and provide public anonymized and aggregated data sets of possible virus spread hotspots location. Similar to MIT Safe Paths, another proposed application is called MiPasa: a global-scale control and communication system that enables detection of COVID-19 carriers and infection hotspots through a fully private information sharing mechanism between multiple stakeholders \cite{mipasa2020}.

Temporary Contact Numbers (TCN) \cite{tnc2020} is a decentralized and privacy-first contact tracing protocol to facilitate information exchange between various contact tracing applications. TCN does not reveal any personal data and allow users' devices to send short-range broadcasts over Bluetooth to nearby devices. This helps users to become informed when they come in contact with an infected person. The TCN protocol also provides a channel for infected persons to report their status. Another protocol similar to TCN is the Private Automated Contact Tracing (PACT) \cite{pact2020}. PACT is a protocol which specifies the broadcast of constantly-changing, anonymous and randomly chosen chirp values on users’ devices to the network. When a user becomes infected, PACT activates risk containment and contact tracing by making public (upon user’s approval) the user's chirp values in the past three weeks.

There are other decentralized applications that have emerged as a result of the current COVID-19 situation in various sectors including: Alipay blockchain for tracking medical supplies \cite{alipay2020}, Public Health Blockchain Consortium (PHBC) on COVID-19 related research \cite{nguyen2020blockchain}, and Hyperchain for tracking donations \cite{hyperchain2020}.

\subsection{Privacy vs utility trade-offs}
The need to contain the spread of COVID-19 through certain measures such as contact tracing and social distancing comes with a necessary requirement to acquire some personal information from citizens. Personal information such as location and health information have been considered important to ensuring that infectious clusters are detected on time so that quarantine measures can be effectively implemented. However, it is also important to respect the privacy rights of citizens as access to such information maybe counter productive during and after the pandemic \cite{cao2020panda}. 

As for the level of threat to privacy based on the type of technology used, we present the results of our evaluation on table \ref{level_threat}.

\begin{table}[htbp] 
\caption{The level of threat to privacy} 
\label{level_threat}
\begin{tabular}{p{2cm}|p{3cm}|p{3cm}|p{3cm} } 
\hline
  &Base station & GPS & Bluetooth  \\ 
\hline 
Threat to the privacy & 
It usually does not constitute a privacy violation concern, since they are involving a large number of people rather than specific personal local information. &  
It is easy to violate the privacy of the user and be rejected by the user. & 
It usually does not constitute a privacy violation concern, since they are involving a large number of people rather than specific personal local information.\\
\hline
The level of the threat& Low &  High& Low\\
\hline
\end{tabular}
\end{table}
\newpage
\subsubsection{Trade-offs}
In table \ref{trade_offs}, we take a look at how various applications handle privacy especially when handling personal information. We also compare how much utility can be provided at varying privacy levels. The \textit{permissibility} (Perm.) refers to what degree of privacy violation can an application have where 0 means privacy rights are fully protected and 1 means a user's privacy rights can be completely ignored. The \textit{utility} metric assesses how much good or original intentions of the application can be fulfilled at varying privacy levels. 

\begin{table}[h!]
\centering
\caption{Utility derived at various privacy levels}
\label{trade_offs}
\begin{tabular}{|p{0.16\linewidth}|p{0.08\linewidth}|p{0.16\linewidth}|p{0.13\linewidth}|p{0.15\linewidth}|p{0.11\linewidth}|}
\hline
\textbf{Application} &  \textbf{Perm.} & \textbf{\textit{Utility \newline @0}} & \textbf{\textit{Utility \newline @100}} &\textbf{\textit{Utility \newline @sys}} & \textbf{Checker}\\
\hline
MIT SafePaths \cite{raskar2020apps} &  0.3 & PDF + OF & (0.7*PDF)    + OF & (0.3 * PDF) + OF & Open\\
\hline
MiPasa \cite{mipasa2020} &  0.67 & PDF + OF & (0.33*PDF) + OF & (0.67 * PDF) + OF & N/A\\
\hline
TNC \cite{tnc2020} &  0.3 & PDF + OF & (0.7*PDF) + OF & (0.3 * PDF) + OF & Open\\
\hline
PACT \cite{pact2020} &  0.3 & PDF + OF & (0.7*PDF) + OF & (0.3 * PDF) + OF & Open\\
\hline
\end{tabular}

PDF = Privacy Dependent Features, OF = Other Features
\end{table}

For example, \textit{utility@0} would mean the utility provided to the user with no privacy guarantee while \textit{utility@100} means utility provided with no potential threats to privacy. The \textit{utility@sys} is the base utility the system can provide by itself whether or not the user consents to use of personal information. The final parameter is the \textit{checker} which is the entity that can verify privacy preservation or violation. Checkers can be open (anyone), collective (multi-party) or sovereign (single entity) checker. We calculate the permissibility and measure utility by providing answers to the following questions:

\begin{enumerate}
    \item Can the application take total control of the user’s privacy? Yes = 30, No = 0 (Utility: How many features depend on its ability to do so?)
    \item If No to (1), does the application collect any personally identifying information? Yes = 10, No = 0 (Utility: How many features depend on this?)
    \item If (2) = Yes, is this collection consent based? Yes = 0, No = 20 
    \item If consent to collect personal data is denied, can the application perform all targeted goals? Yes = 0, No = 5 (Utility: How many can be performed if not all?)
    \item Is the application data decentralized? Yes = 0, No = 10
    \item Is the application’s data storage done locally or  cloud-based? Yes (cloud) = 0(10), No(local) = 10 (0)
    \item Does the data collection process and usage actively involve users or it is done separately? Yes(separate) = 5(10), No(active) = 10(5)
    \item Is the application’s code openly accessible? Yes = 0, No = 10
    \item Can a user request for their data to be completely erased? Yes = 0, No = 10
    \item Is this application verified or backed by an independent or non-governmental organization? Yes = 0, No = 10
\end{enumerate}

\subsection{Adoption phase}
In the first phase of the COVID-19 pandemic, the focus has been on identification, testing and care for infected persons. To facilitate identification of infectious groups, contact tracing applications such as \cite{raskar2020apps,tnc2020,pact2020,chan2020pact} are considered important. In the second phase where social distancing measures are gradually being lifted, it becomes important to answer some key questions: How can we prevent a second outbreak (pandemic prevention - prevent)? What places or activities pose a high risk of exposure to being infected (exposure)? Who is infected/infectious, susceptible or immune to the virus (status)? What research paths are promising or what elements are desirable for a successful discovery (research)? When a vaccine is found, how should vaccination be carried out (equity)?

\begin{table}
\caption{Adoption in Phase-2: How COVID-19 apps may be useful}\label{adopt}
\begin{tabular}{|p{0.28\linewidth}|P{0.1\linewidth}|P{0.12\linewidth}|P{0.08\linewidth}|P{0.11\linewidth}|P{0.08\linewidth}|}
\hline
\textbf{Application} &  \textbf{prevent} & \textbf{exposure} & \textbf{status} &\textbf{research} & \textbf{equity}\\
\hline
MiPasa \cite{mipasa2020} &  \xmark & \cmark & \xmark & \cmark & \xmark\\
\hline
MIT SafePaths \cite{raskar2020apps} &  \cmark &  \cmark &  \cmark & \xmark & \xmark\\
\hline
TNC \cite{tnc2020} &  \cmark &  \cmark &  \cmark & \xmark & \xmark\\
\hline
PACT \cite{pact2020} &  \cmark &  \cmark &  \cmark & \xmark & \xmark\\
\hline
\end{tabular}

\end{table}

In the table \ref{adopt} we identify some COVID-19 applications and their ability to provide answers to questions that will be useful in phase 2.

%
%
%
 \bibliographystyle{unsrt}
 \bibliography{main}
\end{document}